\documentclass{article}

\usepackage{array}
\usepackage{booktabs}
\usepackage{multirow}
\usepackage{algorithm2e}
\usepackage{amsmath}
\usepackage{amssymb}
\usepackage{graphicx}
\usepackage{subscript}
\providecommand{\tabularnewline}{\\}
\usepackage{soul}
\usepackage[dvipsnames]{xcolor}

\usepackage{spconf,amsmath,graphicx}
\usepackage{cite}
\usepackage{amsfonts}\usepackage{amsthm}\usepackage{graphicx}
\usepackage{textcomp}
\usepackage{xcolor}
\usepackage{url}
\usepackage{tablefootnote}
\usepackage{cuted} 

\usepackage{algorithmic}
\usepackage{multirow}
\usepackage{subcaption}
\usepackage{caption}
\usepackage[T1]{fontenc}
\newcommand{\SUB}[1]{\ENSURE \hspace{-0.15in} \textbf{#1}}

\usepackage{adjustbox}

 \usepackage{textcomp}
 \newcommand{\textapprox}{\raisebox{0.5ex}{\texttildelow}}

\title{Federated Self-Supervised Learning for Acoustic
Event Classification
}
\twoauthors
 {\hspace{-0.2in}Meng Feng$^1$\sthanks{The work
was done during Meng’s internship at Amazon.},}
{}
  {\hspace{-0.95in}Chieh-Chi Kao$^2$, Qingming Tang$^2$, Ming Sun$^2$, Viktor Rozgic$^2$, Spyros Matsoukas$^2$, Chao Wang$^2$}
	{\hspace{-2.5in}Massachusetts Institute of Technology$^1$     \hspace{1.75in}    Amazon.com Inc$^2$}

\makeatother
\begin{document}
%
\maketitle
\begin{abstract}
Standard acoustic event classification (AEC) solutions require large-scale
collection of data from client devices for model
optimization. Federated learning (FL) is a compelling framework
that decouples data collection and model training to enhance customer
privacy. In this work, we investigate the feasibility of applying
FL to improve AEC performance while no customer
data can be directly uploaded to the server. We assume no pseudo
labels can be inferred from on-device user inputs, aligning with the typical
use cases of AEC. We adapt self-supervised learning to the
FL framework for on-device continual learning of representations,
and it results in improved performance of the downstream AEC 
classifiers without labeled/pseudo-labeled data available. 
Compared to the baseline w/o FL, the proposed method improves precision up to 20.3\% relatively while maintaining the recall.
Our work differs from prior work in
FL in that our approach does not require user-generated learning targets,
and the data we use is collected from our Beta program and is de-identified, to maximally simulate the
production settings.

\end{abstract}
\section{Introduction}

Acoustic event classification (AEC) is the task of automatically detecting
the occurrence of a set of events within the sound clips recorded
from the target environments. The target events can range from a pre-engineered list such as baby crying and dog barking to event types specified
by the customers themselves. AEC has played an important role in a
wide range of applications in the domain of 
recommendation systems~\cite{cano2005content}
and urban noise analysis~\cite{SONYC}.
It has been conventionally studied with classical speech recognition
techniques \cite{mesaros2010acoustic}\cite{gemmeke2013exemplar}\cite{mesaros2015sound}
and more recently overtaken by deep learning algorithms \cite{cakir2017convolutional}\cite{kao2018r}\cite{jeong2017audio}\cite{wang2019comparison}
thanks to the advancements in machine learning. Recent state-of-the-art
works on AEC commonly need to collect a large set of data on the server to support complex
model optimization routines. The strong coupling between
data and model under the centralized optimization framework exposes privacy risks. 


Federated learning (FL) provides a compelling alternative framework to achieve this goal. 
FL is a distributed learning framework that exploits distributed resources to collaboratively train a machine learning model \cite{liu2021distributed}. 
Thanks to the decoupling of data and model, it is able to keep the sensitive training data locally at the participating devices and never collect them centrally. 
Numerous recent successes \cite{yang2018applied}\cite{hard2018federated}
have shown the viability of applying FL to boost privacy preservation
as well as offer competitive model performance. However, these works
assume access to data annotations directly from user inputs. Unfortunately,
users rarely have any interaction with the
client devices in a typical AEC setting. Consequently, it is difficult to obtain data annotations
directly from customers. In this work, we assume no annotated data
are available besides a small annotated dataset from internal users.
Since the classifier models naïvely trained from this dataset are
likely not able to generalize well for general public users, it is
necessary to take advantage of the customer data locally stored on
client devices for the learned models to generalize to more client
users. 


In this paper, we apply federated learning to improve the performance
of realistic AEC tasks without sending audio to the cloud. The goal is to
improve AEC model precision and generalization to unseen client users
after the deployment of the initial model trained on the small annotated
dataset. We propose a self-supervised federated learning framework
that learns improved representations from the un-labeled audio data
stored on local client devices. Our empirical findings show that improvement
of learned representations after federated learning can lead to improvement
of classification performance even the classifiers are trained on the
same annotated dataset.
Unlike prior work done on public datasets~\cite{saeed2020federated}, we conduct our experiments with internal de-identified data collected from our Beta program.
Our dataset closely resembles the non-independent and identically distributed (IID) distribution of data and devices from highly realistic production settings, which no public datasets for AEC~\cite{gemmeke2017audio}~\cite{mesaros2017dcase} can simulate. 

\section{Related Work}

There has been a great volume of work on learning when labeled data
is scarce. A common class of solutions focuses on learning condensed and generalizable
high-level representations from the surface features such as raw audio
waveforms or spectrograms. 
For example, representations can be learned from autoregressive predictive coding (APC) \cite{chung2019unsupervised}\cite{chung2020generative}\cite{chung2020improved}, 
PASE~\cite{PASE}\cite{PASE_music},
or triplet loss \cite{jansen2018unsupervised}\cite{shor2020towards}.
These self-supervised learning techniques may benefit downstream tasks such as AEC. 
Small encoder models \cite{tagliasacchi2019self}
applicable for mobile devices also share similar findings. Our work
builds on top of \cite{chung2020generative} to extract high-level
features via federated learning.


Federated learning is able to indirectly learn from an increasing
amount of data without sending audio to the cloud as opposed
to being limited to a fixed dataset. Recent work \cite{yang2018applied}
\cite{hard2018federated}\cite{guliani2021training}\cite{cui2021federated}\cite{gao2021end}
shows that FL can output competitive models when the learning targets
are given via supervised learning~methods. However, these works either
assume available training targets from user inputs or perform studies
in a different field. Perhaps the closest approach to ours is \cite{saeed2020federated},
in which federated self-supervised learning is applied to learn representations
from multi-sensor data. Since its users are simulated by randomly
dividing the training set, its experiment cannot simulate the non-IID
distribution of data and devices. In comparison,
we conduct our experiment on internal datasets, where the
partitioning of users and devices is real rather than simulated. We
simulate the assumptions from realistic production settings that none
of our customer data is uploaded or accessible. We use federated
self-supervised learning to improve learned representations, and we
show that the improvement in representation learning translates to
an improvement in event classification performance with no addition of
labeled data. To the best of our knowledge, this work is the first to apply
self-supervised FL to AEC problems on an industrial-level dataset
under realistic assumptions. Our work provides clear evidence of the
viability of FL when data labels are unavailable. 

\begin{figure}[t]
\begin{centering}
\includegraphics[width=0.95\linewidth]{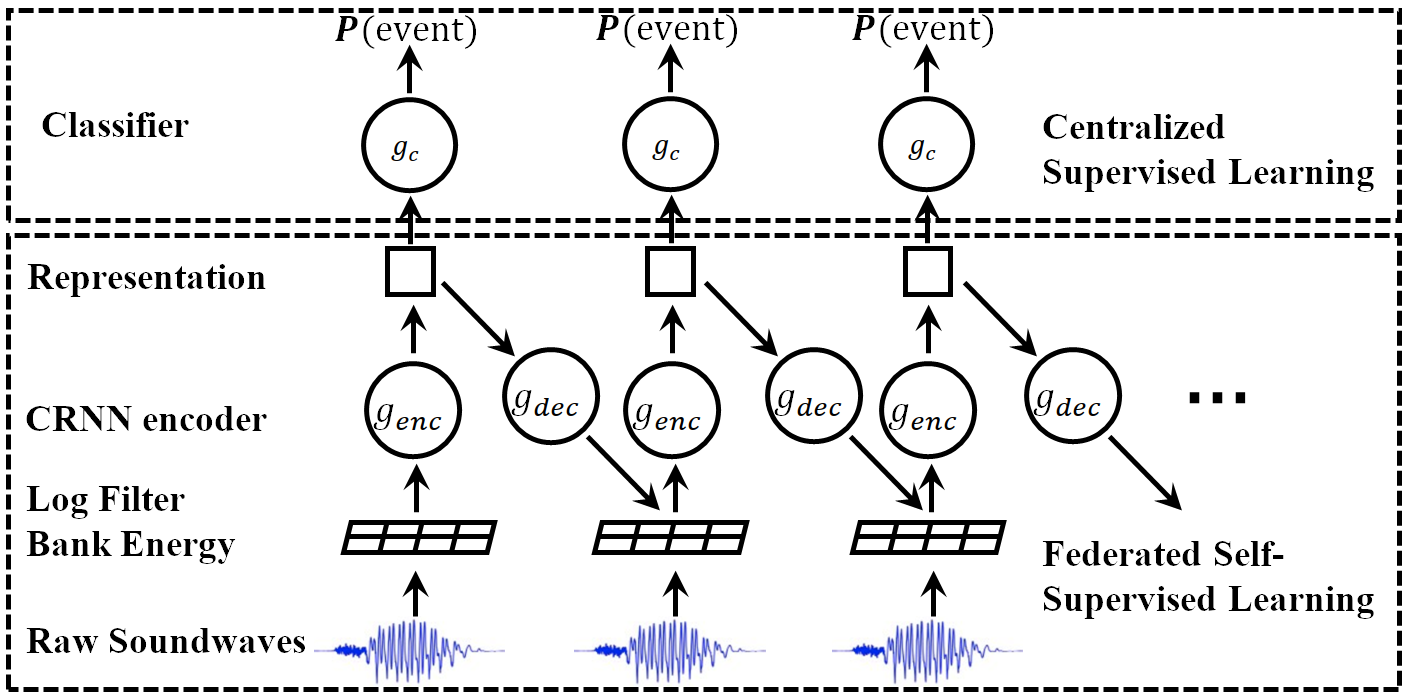}
\par\end{centering}
\caption{Our proposed model architecture. We decouple representation learning
and downstream training of classifiers. We first train a feature encoder
$g_{enc}$ to encode input signals (e.g. LFBE) to latent feature space.
We apply federated learning to expose models to an increasing pool
of users for benefits in the generalization of the learned representations.
We then apply conventional supervised learning on the server using
server data to fine-tune the classifier. 
}
\label{fig:arch}
\end{figure}


\section{Methods}

Given an audio signal $\boldsymbol{x}=(x_{1},x_{2,},...,x_{N})$,
where $N$ is the length of a full utterance, we consider the task
to train a predictor $\boldsymbol{f}$ make a binary prediction $\boldsymbol{z}\in\{0,1\}$
on whether a certain event presents in $\boldsymbol{x}$. We denote $D_{Server}=\{(\boldsymbol{x},\boldsymbol{z})\}$ as the fully annotated dataset on the server and $D_{Client}=\{\boldsymbol{x}\}$
as the unlabeled client dataset stored in the client devices from
customer client devices $S=(S_{1},...,S_{K})$.


As shown in Fig \ref{fig:arch}, we first train an APC model that
consists of an encoder $g_{enc}:x\rightarrow h,x\in\mathbb{R}^{n},h\in\mathbb{R}^{m}$
and a decoder $g_{dec}:h\rightarrow x,h\in\mathbb{R}^{m},x\in\mathbb{R}^{n}$,
where $m,n$ are the dimension of the latent feature vector and post-processed
input (e.g., LFBE) respectively. The decoder then uses the feature
vector as the input to predict a future frame $x_{i+n}$, where $n$
is the number of steps the prediction is ahead of $x_{i}$. We optimize the L1 reconstruction loss
between the predicted sequence $\boldsymbol{y}=(y_{1},y_{2},...,y_{N})$
and the target sequence $\boldsymbol{t}=(x_{1+n},x_{2},...,x_{N+n})$.
$w_{enc}$ and $w_{dec}$ are the parameters of the encoder and decoder
respectively.


\setlength{\textfloatsep}{4pt}

To take advantage of locally stored data $D_{client}$, we apply the
\textsf{\texttt{FederatedAveraging}} algorithm \cite{mcmahan2017communication}.
We first train an initial APC model $\mathcal{M}_{0}$ from an annotated
dataset. This dataset
is assumed to be collected from internal testing participants who
explicitly agree to upload their data to the server. $\mathcal{M}_{0}$
is sent to all client devices and serves as the global starting point
for federated learning at $t=0$. Each of the participating client
devices in a given round of communication accumulates a local dataset
$\mathcal{D}_{k}$, where $k$ is the index of the client device.
The size of the dataset $D_{k}$ may vary from device to device. Each
client device optimizes its local model on its local dataset by running
stochastic gradient descent (SGD) on L1 loss. The model weights of a selected set of client devices
are uploaded back to the server at the end of the communication. The
server aggregates the weights of the models to obtain a new global
model $\mathcal{M}_{1}$. $\mathcal{M}_{1}$ is sent and optimized
on the participating devices in the next round of communication. This
process repeats to incorporate an increasing amount of decentralized data in training the global model $\mathcal{M}$. In essence, after adopting the federated learning framework the loss
function can be written as Eq \ref{eq:fl_loss}, where $n_{k}=|\mathcal{D}_{k}|$ and $n=\sum_{k=1}^{K}n_{k}$.



\vspace{-0.2in}
\begin{flalign}
 & \min_{w_{enc},w_{dec}}L(w_{enc},w_{dec})\nonumber \\
\textnormal{s.t.,\ensuremath{\quad\quad}} & L(w_{enc},w_{dec})=\sum_{k=1}^{K}\frac{n_{k}}{n}l(w_{enc},w_{dec})\label{eq:fl_loss}\\
 & l_{w_{enc},w_{dec}}=\sum_{i=1}^{N-n}|x_{i+n}-y_{i}|\nonumber 
\end{flalign}
Applying federated learning to train $\mathcal{M}$ on client data
can improve the generalization of the learned feature encoders to
customers who are not in the internal testing program. Once $\mathcal{M}$
converges, its parameters are frozen.

To predict event occurrences, we train a classifier $g_{c}:h\rightarrow p,h\in\mathbb{R}^{m},p\in\mathbb{R}$
which takes the input of the encoded feature vectors and outputs the
binary predictions $\boldsymbol{z}\in\{0,1\}$. The classifiers are
trained with a modified binary cross-entropy (BCE) loss seen in Eq \eqref{eq:classifier_loss},

\vspace{-0.2in}
\begin{equation}
l_{c}=-\left[c\cdot z_{n}\cdot\ln p_{n}+(1-z_{n})\ln(1-p_{n})\right]\label{eq:classifier_loss}
\end{equation}

\noindent where $c$ is a positive scaling factor to adjust the loss for positive
samples.  In this paper, we conduct
our experiment on single event classification. However, the same procedures
can be easily extended to multi-class classification problems by adding
a binary classifier for each new class of event following
a one-vs-all paradigm.

\begin{algorithm}[t]
\caption{{\fontfamily{qcr}\selectfont Federated Self-Supervised Federated Learning (FSSL)}. The $K$ clients are indexed by $k$; $B$ is the local minibatch size, $E$ is the number of local epochs, and $\eta$ is the learning rate.} \label{alg:fssa} 
\begin{algorithmic} 
\SUB{Server executes}  
    \STATE initialize APC model weights $w = (w_{enc}$, $w_{dec})$  
    \STATE initialize the classification-layer parameter $w_{c}$  
    \STATE // \textit{stage I: pre-train} 
    \STATE train $w_{0}$ on $\mathcal{D}_{server}$  
    \STATE // \textit{stage II: federated self-supervised learning} 
    \FOR{each round $t=1,2,...$}     
        \STATE $S_t \leftarrow$ (random set of $k$ clients)     
    \FOR{each client $k \in S_t$ 
        \textbf{in parallel}}         
        \STATE $w_{t+1}^k \leftarrow$ ClientUpdate($k, w_t$)     
    \ENDFOR     
    \STATE $w_{t+1} \leftarrow \sum_{k=1}^K \frac{n_k}{n} w_{t+1}^k$ 
    \ENDFOR \STATE // 
    \textit{stage III: train classifier on $\mathcal{D}_{server}$} 
    \STATE Fix $w_{enc}$ 
    \FOR{$\mathcal{B}$ in $\mathcal{D}_{server}$}     
        \STATE $h \leftarrow g_{enc}(\mathcal{B})$     
        \STATE $p' \leftarrow g_{c}(h)$     
        \STATE $w_c \leftarrow w_c - \eta \nabla l_c(w_c, p; \mathcal{B})$ 
    \ENDFOR
\SUB{ClientUpdate($k$, $w$)}: 
    \STATE $\mathcal{B} \leftarrow$ (split $\mathcal{D}_k$ into batches of size $B$)  \FOR{each local epoch $i$ from 1 to $E$}     
        \FOR{batch $b \in \mathcal{B}$}         
            \STATE $w \leftarrow w - \eta \nabla l(w;b)$     
        \ENDFOR 
    \ENDFOR 
\end{algorithmic}
\label{algo:rbac} 
\end{algorithm}



\begin{figure*}[t] 	
\vspace{-0.1 in}
	\centering 	
\begin{subfigure}[b]{0.32\textwidth}
	\centering 	
	\includegraphics[width=\textwidth]{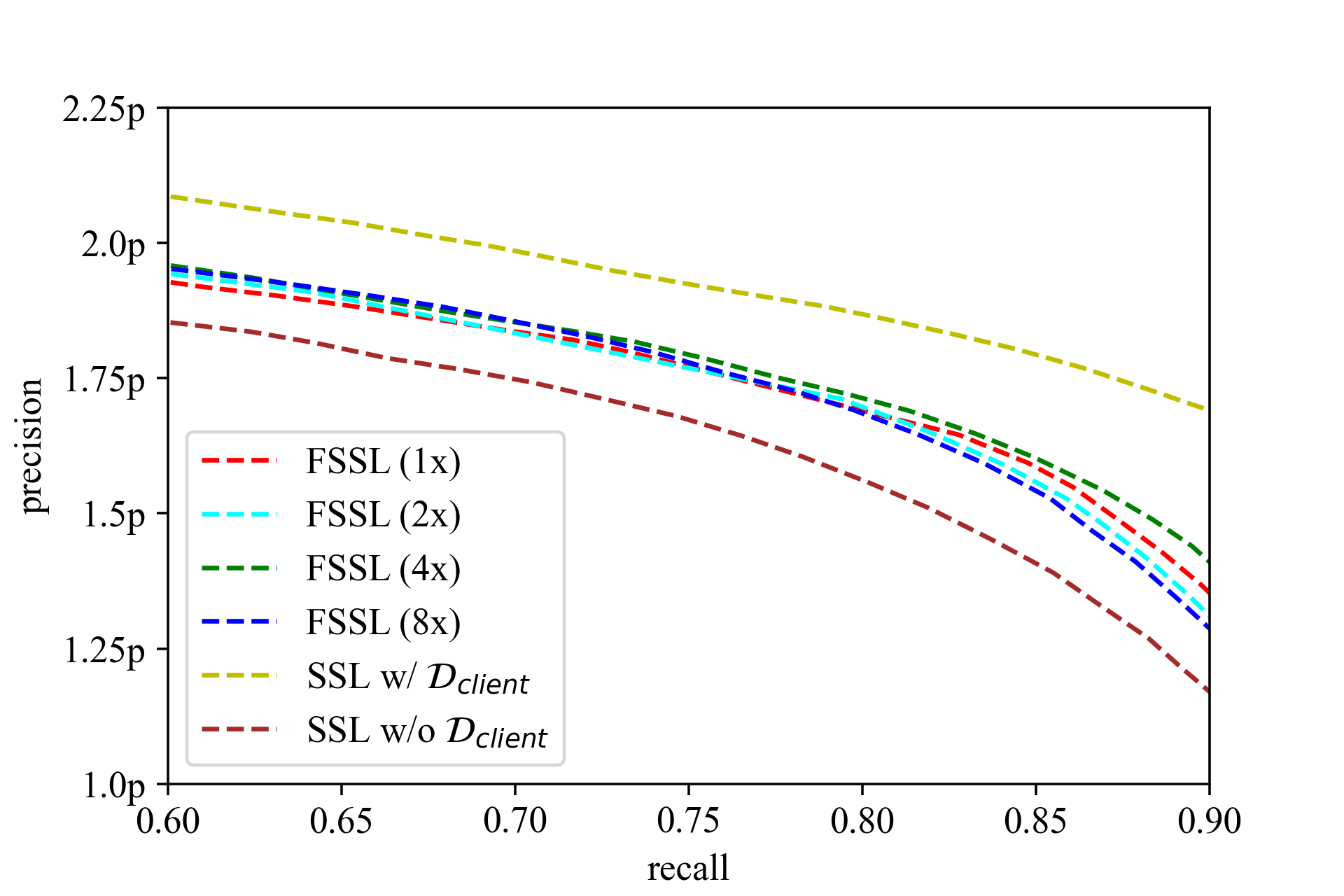}
	\caption{$\mathcal{D_{\mathcal{I}}}$}
	\label{fig:train_all}
\end{subfigure} 
\begin{subfigure}[b]{0.32\textwidth}
	\centering 	
\includegraphics[width=\textwidth]{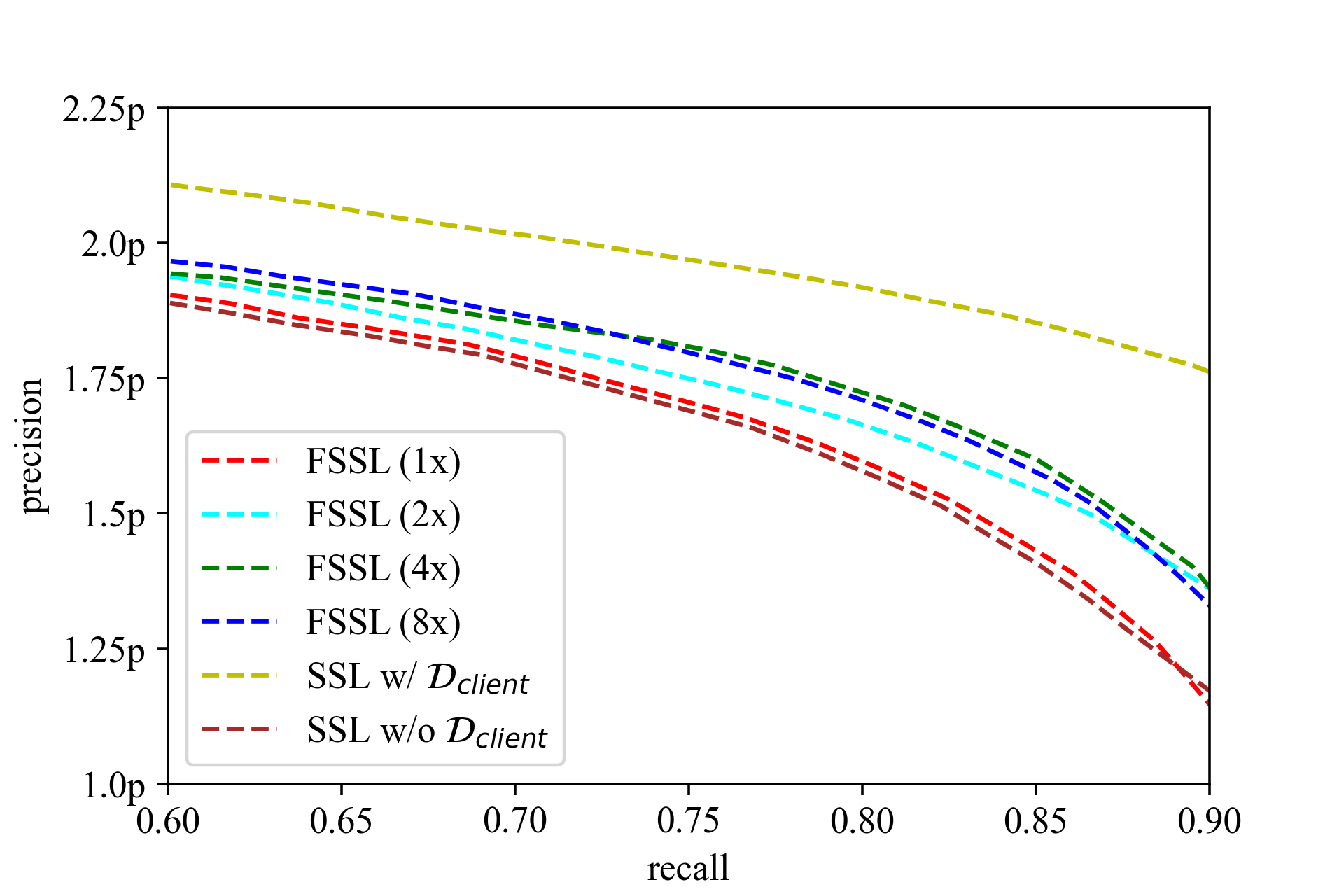}
	\caption{$\mathcal{D_{\mathcal{U}}}$}
	\label{fig:train_unseen_all}
\end{subfigure} 
\begin{subfigure}[b]{0.32\textwidth}
	\centering 	
	\includegraphics[width=\textwidth]{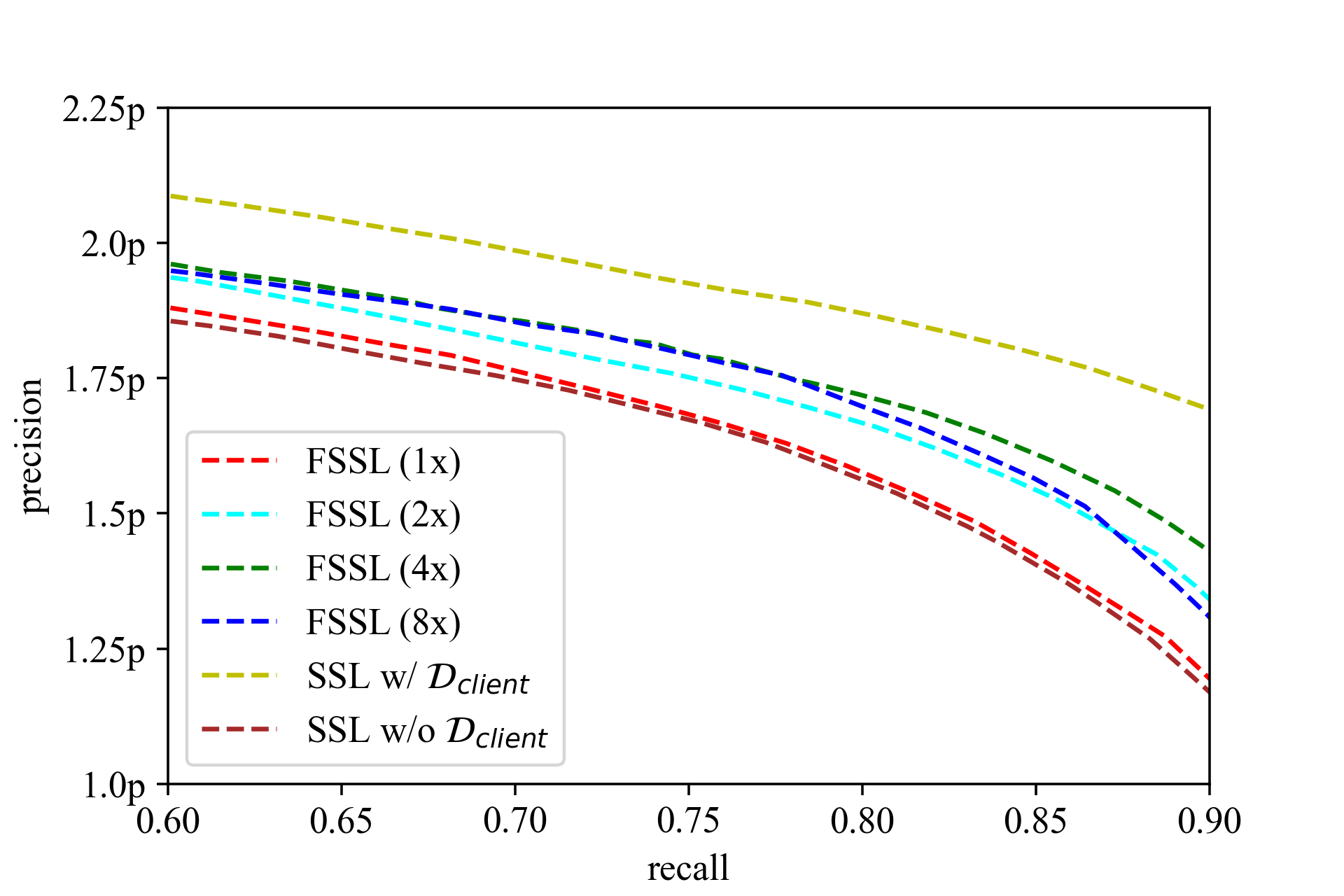}
	\caption{$\mathcal{D_{\mathcal{T}}}$}
	\label{fig:test_all}
\end{subfigure}
    \vspace{-0.1in}
	\caption{
    The effect of the size of the client data $\mathcal{D}_{client}$ on precision-recall curves. $p$ is a constant. For reference, FSSL (1x) model is trained on $\mathcal{D}_{client}$ containing 287,302 utterances. FSSL(2x), FSSL(4x), and FSSL(8x) are trained on augmented $\mathcal{D}_{client}$ by incorporating extra un-labeled data.
	}

	\label{fig:ablation_data}
\end{figure*}

\begin{table*}[t]
\begin{adjustbox}{max width=\textwidth}
\begin{tabular}{ccccccccccccc}
\toprule 
\multirow{3}{*}{\textbf{\scriptsize{}Methods}} & \multicolumn{3}{c}{\textsf{\textbf{\scriptsize{}$\mathcal{D}$}}\textsubscript{\textsf{\textbf{\scriptsize{}$\mathcal{I}$}}}} &  & \multicolumn{4}{c}{\textsf{\textbf{\scriptsize{}$\mathcal{D}$}}\textsubscript{\textsf{\textbf{\scriptsize{}$\mathcal{U}$}}}} & \multicolumn{4}{c}{\textsf{\textbf{\scriptsize{}$\mathcal{D}$}}\textsubscript{\textsf{\textbf{\scriptsize{}$\mathcal{T}$}}}}\tabularnewline
\cmidrule{2-13} \cmidrule{3-13} \cmidrule{4-13} \cmidrule{5-13} \cmidrule{6-13} \cmidrule{7-13} \cmidrule{8-13} \cmidrule{9-13} \cmidrule{10-13} \cmidrule{11-13} \cmidrule{12-13} \cmidrule{13-13} 
 & \multirow{2}{*}{\textbf{\scriptsize{}AUC (\%)}} & \multicolumn{3}{c}{\textbf{\scriptsize{}Precision (\%) at recall $r$}} & \multirow{2}{*}{\textbf{\scriptsize{}AUC (\%)}} & \multicolumn{3}{c}{\textbf{\scriptsize{}Precision (\%) at recall $r$}} & \multirow{2}{*}{\textbf{\scriptsize{}AUC (\%)}} & \multicolumn{3}{c}{\textbf{\scriptsize{}Precision (\%) at recall $r$}}\tabularnewline
\cmidrule{3-5} \cmidrule{4-5} \cmidrule{5-5} \cmidrule{7-9} \cmidrule{8-9} \cmidrule{9-9} \cmidrule{11-13} \cmidrule{12-13} \cmidrule{13-13} 
 &  & {\scriptsize{}$r=0.7$} & {\scriptsize{}$r=0.8$} & {\scriptsize{}$r=0.9$} &  & {\scriptsize{}$r=0.7$} & {\scriptsize{}$r=0.8$} & {\scriptsize{}$r=0.9$} &  & {\scriptsize{}$r=0.7$} & {\scriptsize{}$r=0.8$} & {\scriptsize{}$r=0.9$}\tabularnewline
\midrule
{\scriptsize{}SSL w/o $\mathcal{D}_{client}$} & - & - & - & - & - & - & - & - & - & - & - & -\tabularnewline
{\scriptsize{}FSSL} & 4.55$\uparrow$ & 7.41$\uparrow$ & 10.51$\uparrow$ & 18.56$\uparrow$ & 3.15$\uparrow$ & 2.94$\uparrow$ & 7.56$\uparrow$ & 20.30$\uparrow$ & 4.97$\uparrow$ & 5.72$\uparrow$ & 8.59$\uparrow$ & 18.80$\uparrow$\tabularnewline
{\scriptsize{}SSL w/ $\mathcal{D}_{client}$} & 13.17$\uparrow$ & 14.01$\uparrow$ & 21.60$\uparrow$ & 39.57$\uparrow$ & 16.96$\uparrow$ & 17.71$\uparrow$ & 22.75$\uparrow$ & 52.36$\uparrow$ & 14.87$\uparrow$ & 13.15$\uparrow$ & 23.39$\uparrow$ & 50.83$\uparrow$\tabularnewline
\bottomrule
\end{tabular}
\end{adjustbox}
\caption{Relative benchmark results for the proposed method and baseline models.
Note that we set ``SSL w/o $\mathcal{D}_{client}$'' as the baseline
(100\%) to calculate the relative performance for other two methods
in each column. FSSL is equivalent to the FSSL (4x) in Fig. \ref{fig:ablation_data}. FSSL consistently outperforms SSL w/o  $\mathcal{D}_{client}$ baseline on all data partitions. The improvement, however, is less compared to  SSL w/ $\mathcal{D}_{client}$ when the $\mathcal{D}_{client}$ is directly used to train the classifier.
}

\label{table:benchmark}
\end{table*}

\section{Experiments}

\noindent \textbf{Data} 
The data we use is collected from our Beta program and is de-identified. In our primary study, we use fully annotated data from March to July 2021. It consists of 28,069 unique device numbers (DSNs) and 330,412 audio clips. Each audio clip is 10-second
long and contains information on the timestamp and the DSN of the
source device.
The annotated data from March 2021 are used to simulate the server data $\mathcal{D}_{server}$.
The remaining data from April to July 2021 are used to simulate the client data $\mathcal{D}_{client}$. 
Under this simulation setting, $\mathcal{D}_{server}$ is analogous to participants in the internal testing program, and $\mathcal{D}_{client}$ is analogous to production users without sending audio to the cloud after product launch. Let $\Theta(\mathcal{D})$ be the set of unique DSNs in a dataset $\mathcal{D}$, and we define three subsets of users $\mathcal{I} = \Theta(\mathcal{D}_{server}) \cap \Theta(\mathcal{D}_{client})$, $\mathcal{U} =\Theta(\mathcal{D}_{server})^{C} \cap \Theta(\mathcal{D}_{client})$, and $\mathcal{T} =\Theta(\mathcal{D}_{server})^{C} \cap \Theta(\mathcal{D}_{client})^C$ Intuitively, $\mathcal{I}$ corresponds to the users
who participated in both the internal testing program and post-deployment
product improvement program, $\mathcal{U}$ points to the users who only participated in the post-deployment product improvement program, and $\mathcal{T}$ represents the users who were in neither of the programs. In the result  section, we report our model performances tested on these three partitions $\mathcal{D}_{\mathcal{I}}$, $\mathcal{D}_{\mathcal{U}}$, and $\mathcal{D}_{\mathcal{T}}$ for the  ``dog barking'' event respectively to analyze the impact of applying federated learning on in-distribution and out-of-distribution data samples.
 We assume the device is communicated with the server every 24 hours after running the \textbf{ClientUpdate} routine. 
 After each communication, the data stored on the client devices is cleared due to memory constraints of the client device. 
 In our ablative study, we augment the $\mathcal{D}_{client}$ with unlabeled data uniformly subsampled from the same period. 
 We use the labeled data in $\mathcal{D}_{client}$ to estimate the associated model performances.

\noindent \textbf{Implementation details} We first post-process the
raw audio signals by computing their Log Filter Bank Energy (LFBE)
features with window size 25 ms and hop size of 10 ms. The number
of mel coefficients is 20, which results in a log-mel spectrogram
feature of size $998\times 20$. 
Features are further normalized by
global cepstral mean and variance normalization (CMVN). Our encoder consists of 5 layers of convolutional layers followed by an LSTM layer with 64 units, where the kernels and strides are $[(3,3),(3,3),(3,3),(3,1),(3,1)]$ and $[(2,2),(2,2),(2,1),(2,1),(2,1)]$ respectively. 
Our choice of decoder is a Conv1D
layer that reconstructs the LFBE signals. 
The AEC classifier is made by an additional LSTM layer with hidden size of 96 followed by a dense layer on top of the encoder. A sigmoid function then
maps the dense layer output to $p\in[0,1]$. 

\noindent \textbf{Evaluation Metric} 
We evaluate the performance of
models based on the area under the curve (AUC) and the precision-recall curves
on $\mathcal{D}_{\mathcal{I}}$, $\mathcal{D}_{\mathcal{U}}$, and $\mathcal{D}_{\mathcal{T}}$. We compare
the precisions at the recall values ranging from 0.6 to 0.9 as these
regions are of practical interest for real use cases.


\noindent \textbf{Baseline }We compare our model performance with
two baselines: (1){\it SSL w/o $\mathcal{\mathcal{D}}_{client}$}: classifier trained from pre-trained APC model $\mathcal{M}_{0}$.
This corresponds to the method of directly deploying models trained
from server data without FL,
and (2){\it SSL w/ $\mathcal{\mathcal{D}}_{client}$}: classifier trained with all annotated data $\mathcal{D}=\mathcal{\mathcal{D}}_{server} \cup \mathcal{\mathcal{D}}_{client}$.
This corresponds to the scenario where all customer data is directly
accessible under the centralized training framework, where SSL stands for self-supervised learning. 
Note that (2) is unrealistic for real-world applications but it can be treated as an upper bound for classifiers here.
Both models (1) and (2) consume only the server data
for learning the representations, whereas (2) uses all annotated data
and (1) only uses the server data to train the classifiers.


\vspace{-0.2 in}
\section{Results}
To study the effect of the size of the client dataset $\mathcal{D}_{client}$ used for FL training, we further incorporate extra unlabeled data (\textapprox 2M) collected in the same period as $\mathcal{D}_{client}$ to find the best performing setup.
We vary the size of $\mathcal{D}_{client}$ from 1x to 2x, 4x, and 8x.
Fig.~\ref{fig:ablation_data} shows that further model improvement can be achieved by increasing the size of the client dataset. However, such improvement diminishes when the client dataset is expanded to 8x. 
Therefore, we discuss the performance of the proposed method using FSSL(4x) in the following section.


The results of different models benchmarked on $\mathcal{D}_{\mathcal{I}}$, $\mathcal{D}_{\mathcal{U}}$, and $\mathcal{D}_{\mathcal{T}}$ are shown in Table \ref{table:benchmark}\textsf{. }The proposed
method consistently outperforms the baseline model SSL w/o $\mathcal{D}_{client}$
in all benchmarks, whereas SSL w/ $\mathcal{D}_{client}$ yields the
best model performance among all three models. 
Although the improvement in the overall AUC brought by FL of representations
is relatively small, the improvements at high recall regions (e.g., 0.6-0.9) are significant.
Since high recall regions are of practical production interest, our results show
clear evidence that fine-tuning acoustic event classification model
in the post-deployment stage through continual learning of representations
is feasible. The vastly superior performance from SSL w/ $\mathcal{D}_{client}$
the model indicates that when well-annotated data is accessible to centralized
computing resources, conventional supervised learning is still more
effective in optimizing model weights with respect to fixed learning
targets. However, for particular use cases where sensitive data is
involved and the learning algorithm is not allowed to directly interact
with large datasets on the cloud, centralized learning algorithms
is consequently infeasible. In turn, federated learning may be one
of the few solutions to the task.

\vspace{-0.1 in}
\section{Conclusions}

We show that leveraging self-supervised federated learning to train
AEC models on local client data leads to improvement in model performance
with no extra cost of adding labeled data. Although training representation
encoders is relatively less effective than directly training the classifiers
using centralized training methods, we note that federated learning can
become the deciding factor for certain applications given concerns around consumer data privacy. A seemingly promising direction for future research is to explore and apply more complicated federated learning algorithms within the same framework, although it is not within the scope of this work.
\vfill\pagebreak

\bibliographystyle{IEEEbib}
\footnotesize
\bibliography{main}

\end{document}